\documentclass[proof]{pasj00}

\begin{document}
\SetRunningHead{Nogami et al.}{Two Sun-like Superflare Stars}

\title{Two Sun-like Superflare Stars Rotating as Slow as the
Sun\footnote{Based on data collected at Subaru Telescope, which is
operated by the National Astronomical Observatory of Japan.}}

\author{Daisaku \textsc{Nogami}$^1$, Yuta \textsc{Notsu}$^2$,
  Satoshi \textsc{Honda}$^3$, Hiroyuki \textsc{Maehara}$^4$,
  Shota \textsc{Notsu}$^2$, Takuya \textsc{Shibayama}$^2$,
  and Kazunari \textsc{Shibata}$^1$
}

\affil{$^1$Kwasan and Hida Observatories, Kyoto University, Yamashina, Kyoto 607-8471, Japan}
\email{nogami@kwasan.kyoto-u.ac.jp}

\affil{$^2$Department of Astronomy, Faculty of Science, Kyoto University, Sakyo, Kyoto 606-8502, Japan}

\affil{$^3$Center for Astronomy, University of Hyogo, 407-2, Nishigaichi, Sayo-cho, Sayo, Hyogo 679-5313}

\affil{$^4$Kiso Observatory, Institute of Astronomy, School of Science, The University of Tokyo, 10762-30,
Mitake, Kiso-machi, Kiso-gun, Nagano 397-0101}



\maketitle

\begin{abstract}
We report on the results of high dispersion spectroscopy of two
`superflare stars', KIC 9766237, and KIC 9944137 with Subaru/HDS.
Superflare stars are G-type main sequence stars, but show gigantic
flares compared to the Sun, which have been recently discovered in
the data obtained with the Kepler spacecraft.  Though most of these
stars are thought to have a rotation period shorter than 10 days on
the basis of photometric variabilities, the two targets of the
present paper are estimated to have a rotation period of 21.8 d,
and 25.3 d.  Our spectroscopic results clarified that these stars
have stellar parameters similar to those of the Sun in terms of
the effective temperature, surface gravity, and metallicity.  The
projected rotational velocities derived by us are consistent with
the photometric rotation period, indicating a fairy high
inclination angle.  The average strength of the magnetic field on
the surface of these stars are estimated to be 1-20 G, by using the
absorption line of Ca \textsc{ii} 8542.  We could not detect any
hint of binarity in our spectra,
although more data are needed to firmly rule out the presence of
an unseen low-mass companion.  These results claim that the
spectroscopic properties of these superflare stars are very close
to those of the Sun, and support the hypothesis that the Sun might
cause a superflare.
\end{abstract}

\section{Introduction}

A flare event on the Sun was visually noticed, and recorded by
\citet{car1859} for the first time.  Since then, many solar
flares have been observed in all the wavelengths, and these are
now generally explained by explosive release of the magnetic
energy stored around sunspots (for a review, e.g. \cite{shi11}).
The total energy released in the largest solar flares was estimated
to be of the order of 10$^{32}$ erg (e.g. \cite{pri81, ems12}),
while a possibility has been recently discussed that the most
energetic solar flares caused extreme cosmic-ray events evidenced by
isotopic abundance research of historical relics \citep{miy12, uso12,
mel12, miy13}.  Such strong solar flares could have a severe effect
on the terrestrial environment.

Many `superflares' were discovered in G-type main sequence
stars, namely solar-type stars, in 2012, and 2013 \citep{mae12,
shi13b}, while only nine such events had been reported before it
\citep{sch00}.\footnote{Some of these nine flares seem doubtful.
Gmb 1830 (=HR 4550) have a flare star companion, CF UMa (see
\cite{rei91}).  The superflare visually observed on S For is
suggested to be due to mis-identification (see \cite{pay52, ash59}).
In addition, $\kappa$ Cet and $\pi^1$ UMa rotate with a period of
9.24 d, and 4.89 d, respectively, and are rather young
(age $<$ 1 Gy) \citep{lin12}.}  Superflares are eruptive events
having an energy 10 to 10$^6$ times larger than that of the largest
solar flare recorded so far.  Such events have been found in many
T~Tau stars, RS CVn-type binaries, and dMe stars (e.g. \cite{shi02}).
Before the advent of high-precision space photometry, however, it had
been difficult to detect a superflare in white light in solar-type
stars, due to contrast reasons.  Indeed, even the largest solar
flares could increase the total luminosity of the Sun by only 0.03\%
even at the peak of the event (e.g. \cite{kop05}).  This situation has
been recently changed by the Kepler spacecraft which observed a sky
region between Cygnus and Lyra, monitoring over 160 thousand stars
with a cadence of about 30 minutes, and a quite high accuracy
exceeding 0.01\% for moderately bright stars \citep{koc10}.  Kepler
data have been then used for the stellar flare reseach on cool stars
(e.g. \cite{wal11}), and on hot stars \citep{bal12}.  We hereafter
call those solar-type stars showing superflares `superflare stars'.

The analyses of the Kepler data by \citet{mae12} and \citet{shi13b}
revealed that the relation between the occurrence frequency of the
flare and the flare energy in the Sun can be roughly extended for
the flare energy up to $\sim 10^{36}$ erg on superflare stars.
\citet{shi13b} also estimated that a superflare with an energy of
10$^{34-35}$ erg occurs once in 800-5,000 years in Sun-like stars,
i.e. main-sequence stars with the effective temperature in the range
5600--6000 K, and the rotational period longer than 10 days (see
also \cite{shi13a}).  This rotation period is, however, estimated
from the brightness modulation, assuming that it is due to the
rotation of the star, whose photosphere is covered by large
starspots (see \cite{not13b}).  The rotation period then should be
spectroscopically confirmed.  In addition, the stellar parameters,
and binarity of the superflare stars should be checked for the
discussion whether the Sun can really give rise to superflares, or
not.

\begin{figure}
  \begin{center}
    \FigureFile(80mm,115mm){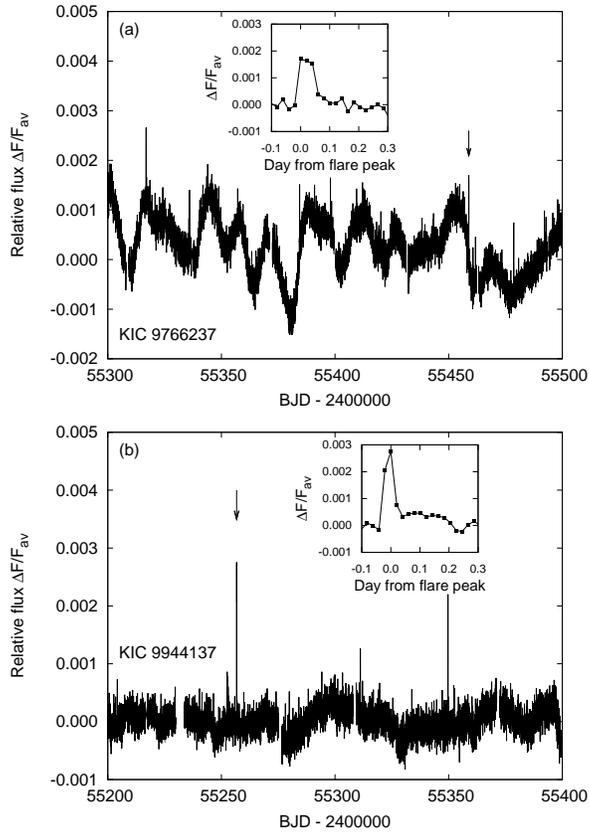}
  \end{center}
  \caption{(a) Typical light curve of KIC 9766237 drawn with the
    long cadence Kepler data.  The Y axis represents the relative
    flux normalized by the average flux ($(F-F_{\rm av})/F_{\rm av}$).
    Quasi-periodic modulations with a timescale of about twenty days
    are seen.  A tick mark points out the superflare automatically
    detected by the method described in \citet{mae12} and \citet{shi13b}.
    The inset figure shows an enlarged light curve around the ticked
    superflare.  The amplitude is about 0.17 \%, and the
    duration is about 0.1 days.  Though many `spikes' other than
    that superflare are seen, all of them consist of only one point,
    and are not judged as a superflare with our criteria.  (b) The
    same as (a), but of KIC  9944137.  The amplitude of the superflare
    in the inset figure is about 0.28 \%, and the duration exceeds 0.2
    days.}
    \label{fig:lc}
\end{figure}

We have then performed high-dispersion spectroscopy of superflare
stars in our database.  In this Letter, we report on the first
results of the analysis of high-resolution spectra of two G-type
stars, namely KIC 9766237 and KIC 9944137, whose rotation period was
estimated to be 21.8, and 25.3 days, respectively \citep{shi13b}.
Figure \ref{fig:lc} shows an example of the Kepler light curve of
these targets, around the only one flare of each star automatically
detected by the method described by \citet{mae12}.  Quasi-periodic
modulations with an amplitude of $\sim$0.1 \%, other than the
superflare, are clearly present.  The total energy released during
the superflares marked in figure \ref{fig:lc} is estimated to be
$\sim10^{34}$ erg, and the duration is 0.1-0.2 days.  Table
\ref{tab:param} summarizes the effective temperature, surface
gravity, and metallicity of KIC 9766237, and KIC 9944137, listed in
the Kepler Input Catalog \citep{bro11}.  Table \ref{tab:param} also
lists the data of 18 Sco, the best investigated solar twin (e.g.
\cite{por97, sou04}), for comparison.

The details of our observation of the targets are summarized in
section 2.  Section 3 describes the observational results, and
brief discussion.

\section{Observations}

The observation of KIC 9944137, and 18 Sco was carried out on 2013
June 23, with the High Dispersion Spectrograph (HDS; \cite{nog02})
attached to the 8.2-m Subaru telescope.  KIC 9766237 was observed
with the same instrument on 2013 June 24.  The total exposure time
for KIC 9766237, KIC 9944137, and 18 Sco was 4800 sec (1200 sec
$\times$ 4 sequential images), 4200 sec (1400 sec $\times$ 3
sequential images), and 20 sec (10 sec $\times$ 2 sequential images),
respectively.  The spectral resolution was $R\sim80,000$.  The
wavelength coverage was 6,100-8,820 \AA, which includes the
chromospheric-activity sensitive lines of H$\alpha$, and Ca \textsc{ii}
IR triplet 8498, 8542, 8662.  The 2$\times$2 on-chip binning mode was
adopted.  We used the image slicer \#2 \citep{taj12}.  We reduced the
Echelle-image data to 1-dimensional spectra in the standard way with
IRAF.\footnote{IRAF is distributed by the National Optical Astronomy
Observatory, which is operated by the Association of Universities for
Research in Astronomy (AURA) under cooperative agreement with the
National Science Foundation.}  The signal-to-noise ratio (S/N) is
S/N=60$\sim$70 around Ca II 8542 for KIC 9766237, KIC 9944137,
and S/N$>$200 for 18 Sco.

The radial velocity of KIC 9766237, KIC 9944137, and 18 Sco was
estimated (see next Section) to be $-$56.1($\pm$0.3),
$-$31.8($\pm$0.3), and 11.9($\pm$0.3) km s$^{-1}$, respectively, by
using the photospheric absorption lines.  Per each star, all the
sequential spectra were combined into a single co-added spectrum,
and the correction for the radial velocities was perfomed before
the following analyses.

\section{Results and Discussion}

\begin{figure}
  \begin{center}
    \FigureFile(80mm,115mm){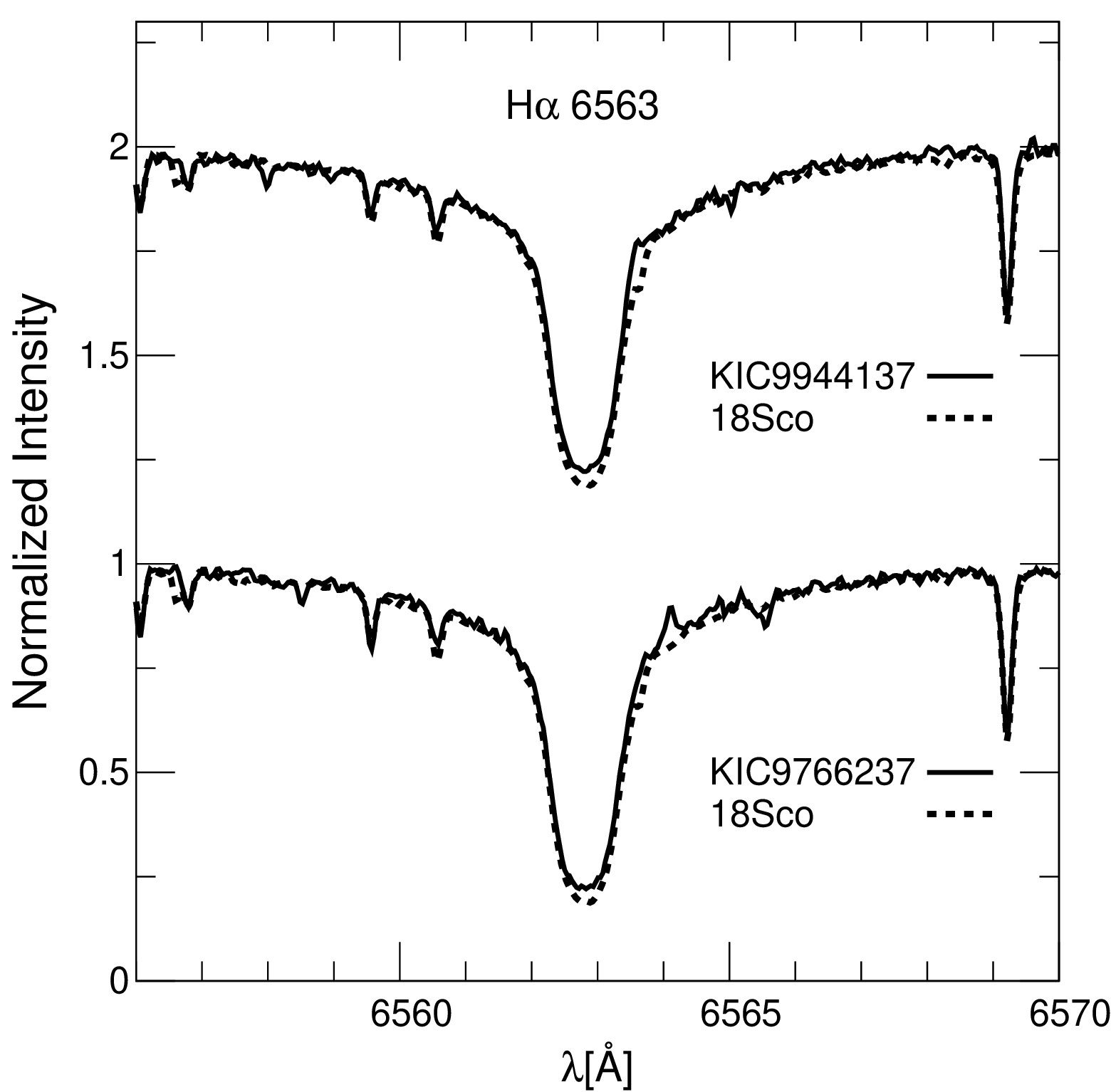}
  \end{center}
  \caption{H$\alpha$ absorption line of KIC 9766237, and KIC 9944137
    (solid lines), and 18 Sco (dashed lines).  The H$\alpha$ depth of
    these targets is slightly smaller than that of the comparison
    solar-twin star, 18 Sco.  The spectrum of KIC 9944137 accompanying
    that of 18 Sco is shifted by +1.0 along the Y axis, for clarity.}
  \label{fig:Ha}
\end{figure}

\begin{figure}
  \begin{center}
    \FigureFile(80mm,115mm){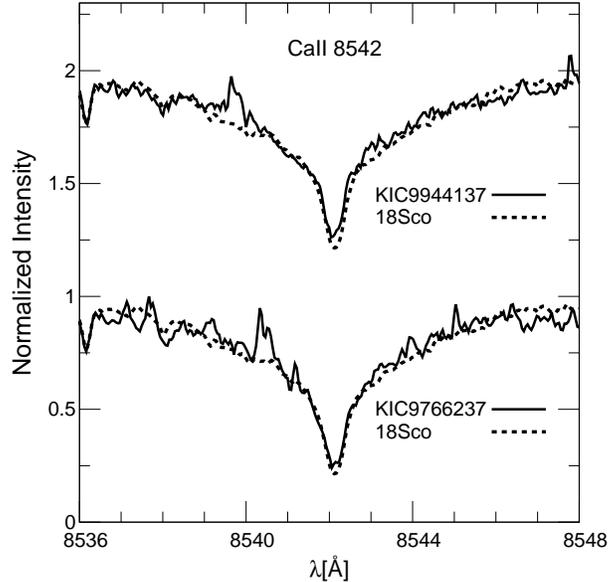}
  \end{center}
  \caption{Absorption profile of Ca \textsc{ii} 8542, one of Ca \textsc{ii}
    IR triplet, of KIC 9766237, and KIC 9944137 (solid lines), and
    18 Sco (dashed lines).  This line of both targets is slightly shallower
    than that of the comparison solar-twin star, 18 Sco.  The spectrum of
    KIC 9944137 accompanying that of 18 Sco is shifted by +1.0 along the
    Y axis, for clarity.}
  \label{fig:Ca}
\end{figure}

Figures \ref{fig:Ha}, and \ref{fig:Ca} show absorption lines of H$\alpha$,
and Ca \textsc{ii} 8542.  They are good indicators of the chromospheric
activity (e.g. \cite{fra10, tak10}).  Both lines of both targets are
slightly shallower than those of the solar twin star, 18 Sco, that has
the same activity level of the quiet Sun on the basis of the Mt Wilson
Ca II S index (e.g. \cite{dun91}).  The index $r_0(8542)$, the flux
normalized by the continuum level at the line core of Ca \textsc{ii}
8542 is 0.26 for KIC 9766237, 0.27 for KIC 9944137, and 0.21 for 18 Sco.
These index values and effective temperature of $T_{\rm eff} =5606
{\rm K}$ for KIC 9766237, and $T_{\rm eff} =5666 {\rm K}$ for
KIC 9944137 (derived later) suggest that both targets are in the region
dominated by quiescent dwarf stars in figure 6 in \citet{not13a}, though
the shallowness of H$\alpha$ and Ca II, compared with those of 18 Sco,
indicates a little higher atmospheric activity level of both stars.

The average magnetic field can be roughly estimated, by using the
emission core flux of Ca \textsc{ii}.  The relation between the emission
core flux of the Ca \textsc{ii} K line and the average magnetic field was
first derived for cool stars by \citet{sch89}.  Notsu et al. (2014, in
preparation) have extended this relation  to the emission core flux of
the Ca \textsc{ii} 8542 and the average magnetic field.  The average
strength of the magnetic field of KIC 9766237, and KIC 9944137 is estimated
to be 1-20 Gauss, by using this method.  This value is of the same order
as, or one order of magnitude higher than that of the Sun ($\sim$2 G).

\begin{figure}
  \begin{center}
    \FigureFile(80mm,115mm){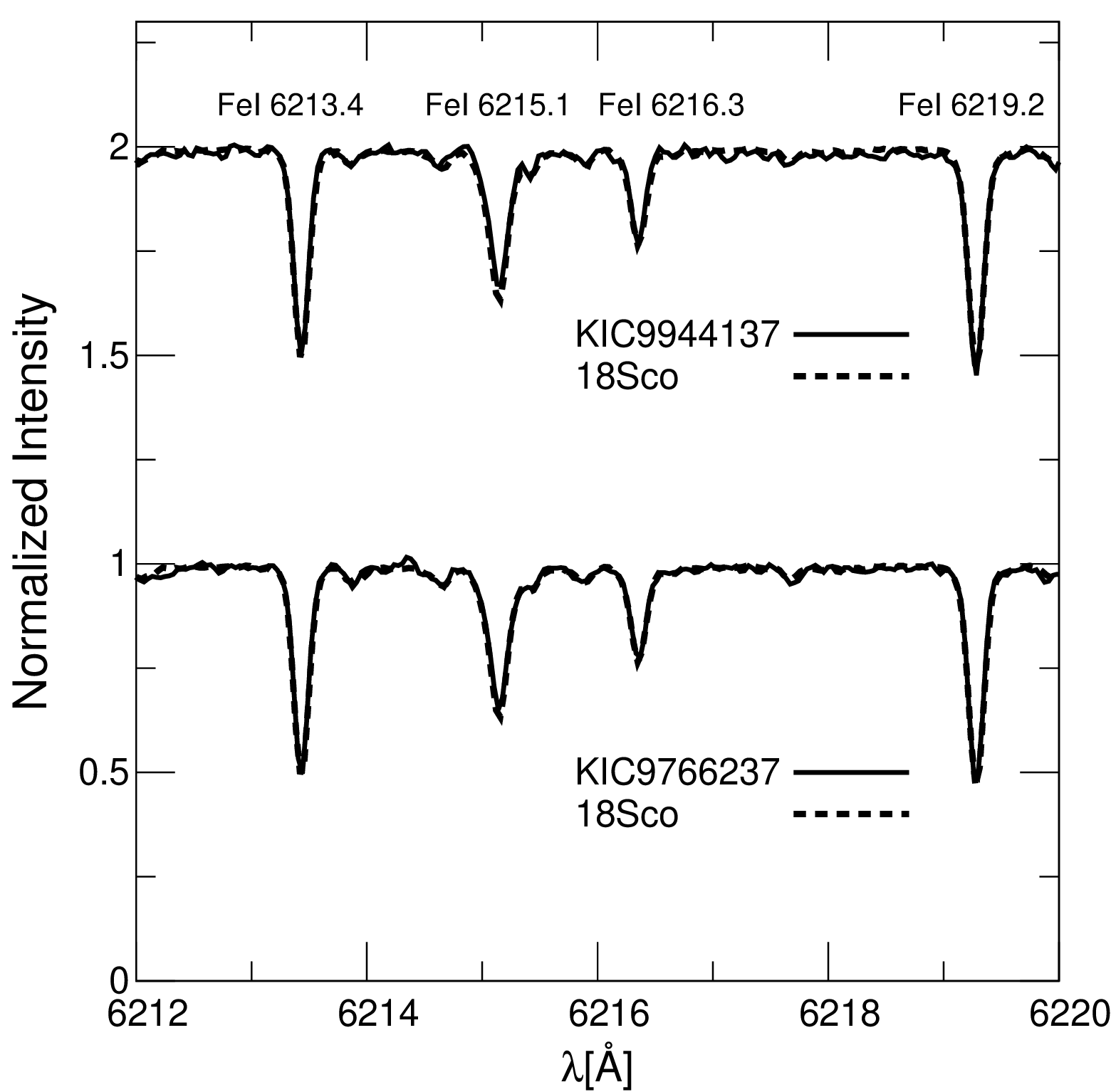}
  \end{center}
  \caption{Example of photospheric absorption lines, including Fe I 6213,
    6215, 6216, and 6219, of the target stars, and 18 Sco, corrected for
    the radial velocity.  The profiles of these lines are in good agreement
    of that of 18 Sco, suggesting that both of the targets are single
    stars.  The projected rotational velocity of KIC 9766237, KIC 9944137,
    and 18 Sco is estimated to be 2.1($\pm$0.3), 1.9($\pm$0.3), and
    2.2($\pm$0.3) km s$^{-1}$, respectively.  The spectrum of KIC 9944137
    accompanying that of 18 Sco is shifted by +1.0 along the Y axis, for
    clarity.}
  \label{fig:Fe}
\end{figure}

An example of the photospheric absorption lines is displayed in figure
\ref{fig:Fe}.  The profile of these photospheric lines are in good
agreement with that of 18 Sco.  This fact suggests that both of KIC
9766237, and KIC 9944137 are single stars, or at most single-lined
binaries.

\begin{table*}
  \caption{Stellar parameters of KIC 9766237, KIC 9944137, and 18 Sco}
  \label{tab:param}
  \begin{center}
    \begin{tabular}{cccccccc}
      \hline\hline
      Name & Sp.   & $T_{\rm eff}$ & $\log g$ & [Fe/H] & $v \sin i$    & 
              A(Li) & Reference$^{\dagger}$ \\
           & Type  &  [K]          & [cgs]    & & [km s$^{-1}$] &  \\
      \hline
      KIC 9766237 & & 5674 & 4.557 & $-$0.147 &  & & KIC \\
                  & & 5606($\pm$40) & 4.25($\pm$0.11) & $-$0.16($\pm$0.04) &
              2.1($\pm$0.3) & $<$1.0 & this study \\
      KIC 9944137 & & 5725 & 4.619 & $-$0.064 &  & & KIC \\
                  & & 5666($\pm$35) & 4.46($\pm$0.09) & $-$0.10($\pm$0.03) &
              1.9($\pm$0.3) & $<$1.0 & this study \\
      18 Sco & G2V & 5816 & 4.45 & 0.04 & 2.1 & 1.63 & 1,2,3 \\
             &     & 5824($\pm$15) & 4.50($\pm$0.04) & 0.07($\pm$0.02) &
              2.4($\pm$0.3) & 1.6($\pm$0.2) & this study \\
      \hline
    \multicolumn{8}{l}{$^{\dagger}$1: \citet{das12}, 2: \citet{pet08},
        3: \citet{mel07}}\\
    \end{tabular}
  \end{center}
\end{table*}

An analysis of the co-added spectrum of KIC 9744237, KIC 9944137, and
18 Sco yields the stellar parameters of the effective temperature
$T_{\rm eff}$, surface gravity $\log g$, projected rotational velocity
$v \sin i$, metallicity [Fe/H], and lithium abundance A(Li)
(table \ref{tab:param}).  The details of this analysis method were
described by \citet{not13a}.

While the values of $T_{\rm eff}$, $\log g$, and [Fe/H] we derived are
rather different than those in the Kepler Input Catalog, spectroscopic
determination of the stellar parameters are known to be more accurate
than the photometric method adopted in the Kepler Input Catalog (e.g.
\cite{mol10, pin12}).  Comparison of our 18 Sco data with those in the
literature supports the accuracy of our analysis.  

The stellar parameters of KIC 9766237, and KIC 9944137 derived here
indicate that both stars have the surface gravity and metallicity
fairly close to those of the Sun, although the effective temperature
of both stars is a little lower ($\sim$100 K) than that of the Sun.

\begin{figure}
  \begin{center}
    \FigureFile(80mm,114mm){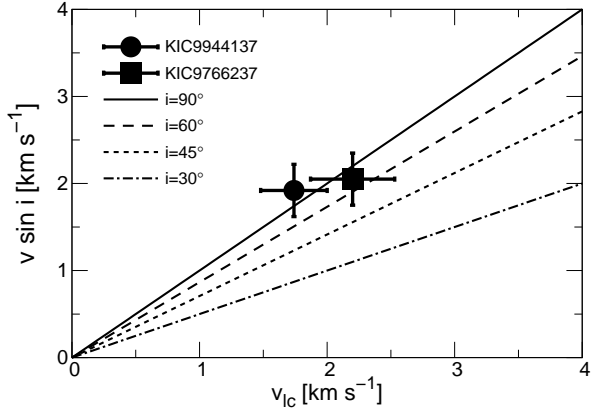}
  \end{center}
  \caption{Diagram of the projected rotational velocity vs. rotational
    velocity estimated from the light curve.  The inclination angle
    is suggested to be fairly high, 90$^{+0}_{-25}$ deg for KIC
    9766237, and 75$^{+15}_{-22}$ deg for KIC 9944137.  The horizontal
    error bar is estimated by taking into account the error of the
    period analysis ($\sim$10 \%), and the error of the radius
    estimation ($\sim$10 \%), and the vertical error bar is estimated
    on an assumption that the error is dominated by the error of the
    estimation of the macroturbulence ($\sim$15 \%; see \cite{hir12}).}
    \label{fig:vsini}
\end{figure}

The projected rotational velocity of 2.1($\pm$0.3) km s$^{-1}$ of KIC
9766237, and 1.9($\pm$0.3) km s$^{-1}$ of KIC 9944137 is consistent
with the photometrically estimated rotation period of 21.8, and 25.3
days, respectively, indicating that the inclination angle $i$ is fairly
high, $i=90^{+0}_{-25}$ deg for KIC 9766237, and $i=75^{+15}_{-22}$ deg
for KIC 9944137 (figure \ref{fig:vsini}).  The method to derive the
rotational velocity on the basis of the light curve was descibed also by
\citet{not13a}, while the data of the isochrone used here was replaced
by that in \citet{bre12}.

\begin{figure}
  \begin{center}
    \FigureFile(80mm,114mm){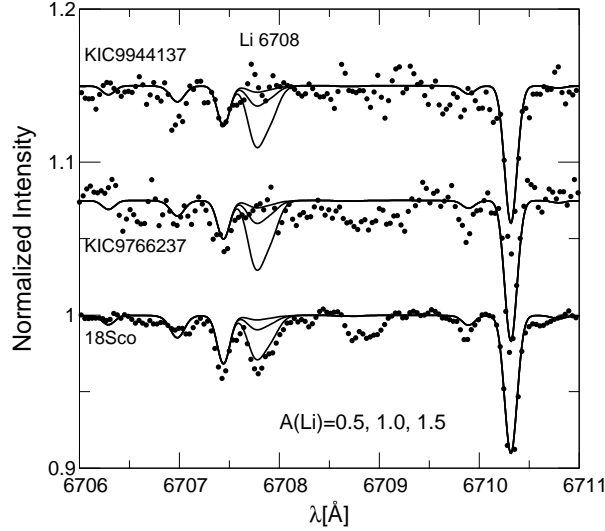}
  \end{center}
  \caption{Li \textsc{i} 6708 absorption line of KIC 9766237, KIC 9944137,
    and 18 Sco.  The solid lines represent synthetic spectra calculated
    for the stellar parameters listed in table \ref{tab:param}, and
    different A(Li) values.  The Li \textsc{i} 6708 line basically become
    shallower with a smaller Li abundance.  The A(Li) value of both
    targets is estimated to be A(Li) $<$ 1.0, although that of 18 Sco is
    estimated to be A(Li)=1.6($\pm$0.2), in good agreement with the values
    in the literature.  The spectrum of KIC 9766237, and KIC 9944137 is
    shifted by +0.075, and +0.15, respectively, along the Y axis, for
    clarity.}
  \label{fig:Li}
\end{figure}

The spectra in the region around 6700 \AA\  are shown in figure
\ref{fig:Li}.  The Li I 6708 \AA\ line is detected neither in the
spectrum of KIC 9766237 nor in that of KIC 9944137, while it is clearly
visible for 18 Sco.  The Li abundance A(Li) is estimated to be smaller
than A(Li)=1.0, on the basis of figure \ref{fig:Li}.  The upper limit
of A(Li) is obtained in another way.  Application of the method
described in \citet{tak05} using the averaged FWHM of weak Fe lines,
and signal-to-noise ratio, to our spectra gives the upper limit of
the equivalent width of this Li absorption EW(Li), EW(Li) $<$ 4.5 m\AA\
for both of our targets.  The upper limit of A(Li) is estimated to be
A(Li) = 1.37 for KIC 9766237, and A(Li) = 1.43 for KIC 9944137, by
calculations of the chemical abundances on the assumption of the local
thermal equilibrium, atmospheric parameters obtained above, and that
upper limit of EW(Li).  We here adopt the A(Li) upper limit of A(Li) =
1.0, that is, A(Li) $<$ 1.0 for our targets.  This value is the same as,
or less than that of the Sun [A(Li)=0.92; \cite{tak10}], and is
relatively small in the range of the effective temperature around 5700 K
\citep{tak07}.  This fact suggests that these stars are not young (e.g.
\cite{dov14}), and both stars have an age greater than 3-4 Gy, unless an
anomalous lithium depletion has occurred.  This is consistent with that
both stars have a long rotation period, while their activity level seems
to be higher than that of the Sun, based on the chromospheric data and
on the lower limit for the relative spot areas of about 0.1 \%, which
is close to the largest sunspots groups observed (0.2 \%).

We here briefly discuss whether superflares can be explained by the
magnetic energy stored on the surface of KIC 9766237, and KIC 9944137.
The average magnetic field of 1-20 G is deduced above, and the total
energy of the superflare of our targets is 10$^{34}$ erg.  On the
assumption that the radiated flare energy is a fraction [$f$;
$f=0.1$ is assumed in \citet{shi13a}] of the magnetic energy
stored around the spot area, the flare energy ($E_{\rm flare}$) is
expressed as below (see \cite{shi13a}):
\begin{equation}
E_{\rm flare} \sim f \frac{B^2}{8\pi}L^{3}
              \sim f \frac{B^2}{8\pi}(a\pi R_*^2)^{3/2}.
\end{equation}
In this equation, the length of the magnetic structure
responsible for the flare, $L$, has been assumed to be equal to
the size of the spotted region, i.e. $L=\sqrt{a*\pi*R_*^2}$, where
$a$ is the area of the spot in units of the stellar disk area.
This equation is transformed by replacing the energy
with 10$^{34}$ erg, as below:
\begin{equation}
a \sim 0.01 \Biggl( \frac{f}{0.1} \Biggr) ^{-\frac{2}{3}}
    \Biggl( \frac{R_*}{\RO} \Biggr)^{-2}
    \Biggl( \frac{B}{10^3\ {\rm G}}  \Biggr)^{-\frac{4}{3}}.
\label{eq:spotarea}
\end{equation}
If the strength of the magnetic field in the spot area is 20 kG, the
fraction $a$ is deduced to be of the order of 0.001 with equation
\ref{eq:spotarea}.  In this case, the average magnetic field results
in being $\sim$20 G, which is in the range derived above.  The 20-kG
magnetic field, however, should be too strong, compared to that of
the Sun, and physically unrealistic, since the magnetic pressure
of the spot area should be up to the same order of the gas pressure
in the photosphere.  If the strength of the magnetic field in the
spot area is 1 kG, the fraction $a$ is deduced to be of the order of
0.01.  In this case, the average magnetic field is of the order of
10 G, which is also in the range derived above.  The fraction a=0.01
is, however, an order of magnitude larger than that evaluated from
the light curve above.  This implies the existence of a large polar
spot which are always visible, and
does not effectively contribute to the observed variability (see e.g.
\cite{hus07}).  More observations are needed for direct measurement
of the spot group area, and the magnetic field strength of the spot
group area.

In conclusion, although KIC 9766237, and KIC 9944137 are superflare
stars having showed a superflare with an emitted energy of 10$^{34}$
erg, these stars turned out to be a G-type main sequence star
with stellar parameters ($T_{\rm eff}$, $\log g$, and [Fe/H])
close to those of the Sun, by our high dispersion spectroscopy.
Their slow rotation period of 21.8 days (KIC 9766237), and 25.3
days (KIC9944137), and very low lithium abundance of A(Li)$<$1.0
indicate that these stars are not young.  The projected rotational
velocity derived here is consistent with the photometric rotation
period, and these data indicate that the inclination angles are
high (70-90 deg).  Line profiles of the photospheric absorption
lines are in good agreement with those of 18 Sco, suggesting that
both of the targets are single stars, though a low-mass companion
can not be excluded without additional observations.  The average
strength of the magnetic field on KIC 9766237, and KIC 9944137 is
estimated to be 1-20 G, by the depth of Ca II 8542.  These
arguments support the hypothesis that the Sun might cause
superflares, though the upper limit of this value (20 G) is one
order of magnitude larger than that of the Sun.  More detailed
observations will be needed to clarify the magnetic field on the
superflare stars.

\bigskip

The authors sincerely thank the referee, Antonio Frasca for his
very useful and constructive comments.  This study is based on
observational data collected at Subaru Telescope, which is
operated by the National Astronomical Observatory of Japan.  We
are grateful to Akito Tajitsu and other staffs of the Subaru
Telescope for making large contributions in carrying out our
observation.  Thanks are also to Youichi Takeda for his many
useful advices on the analysis of our Subaru data, and for his
opening the TGVIT, and SPTOOL programs into public.  We are
indebted to Taichi Kato for his comments on S For.  Kepler was
selected as the tenth Discovery mission.  Funding for this
mission is provided by the NASA Science Mission Directorate.
The Kepler data presented in this paper were obtained from the
Multimission Archive at STScI. This work was supported by the
Grant-in-Aid from the Ministry of Education, Culture, Sports,
Science and Technology of Japan (No. 25287039).


\end{document}